# Strain Solitons and Topological Defects in Bilayer Graphene

Jonathan S. Alden[a], Adam W. Tsen[a], Pinshane Y. Huang[a], Robert Hovden[a], Lola Brown[b], Jiwoong Park[b,c], David A. Muller[a,c], and Paul L. McEuen[c,d*]

[a]School of Applied and Engineering Physics, [b]Department of Chemistry and Chemical Biology, [c]Kavli Institute at Cornell for Nanoscale Science, [d]Laboratory of Atomic and Solid State Physics, Cornell University, Ithaca, NY 14853

* corresponding author, plm23@cornell.edu

**Spontaneous symmetry-breaking, where the ground state of a system has lower symmetry than the underlying Hamiltonian, is ubiquitous in physics. It leads to multiply-degenerate ground states, each with a different "broken" symmetry labeled by an order parameter. The variation of this order parameter in space leads to soliton-like features at the boundaries of different broken-symmetry regions and also to topological point defects. Bilayer graphene is a fascinating realization of this physics, with an order parameter given by its interlayer stacking coordinate. Bilayer graphene has been a subject of intense study because in the presence of a perpendicular electric field, a band gap appears in its electronic spectrum[1-3] through a mechanism that is intimately tied to its broken symmetry. Theorists have further proposed that novel electronic states exist at the boundaries between broken-symmetry stacking domains.[4-5] However, very little is known about the structural properties of these boundaries. Here we use electron microscopy to measure with nanoscale and atomic resolution the widths, motion, and topological structure of soliton boundaries and topological defects in bilayer graphene. We find that each soliton consists of an atomic-scale registry shift between the two graphene layers occurring over 6-11 nm. We infer the minimal energy barrier to interlayer translation and observe soliton motion during in-situ heating above 1000 °C. The abundance of these structures across a variety samples, as well as their unusual properties, suggests that they will have substantial effects on the electronic and mechanical properties of bilayer graphene.**

Spontaneous symmetry-breaking occurs in systems ranging from magnetism in solids to the Higgs mechanism in high energy physics. In the case of a magnet, the spins locally align, creating a magnetization that plays the role of the order parameter. However, the global orientation of the magnetization can be in one of many directions, determined, for example, by the crystal axes. Locally, the system "spontaneously" chooses one such direction based on external constraints or history. Different local regions can have different orientations, and the boundary between adjacent regions is called a domain wall. Mathematically, this boundary takes the form of a soliton that is finite in width but free to move. Other, more complex topological structures are also possible.

The stacking of two graphene sheets exhibits analogous physics. Figure 1a shows the energy of bilayer graphene as a function of the relative in-plane displacement $u$ between the two graphene sheets.[6] The energy as a function of $u$ is maximal in the high-symmetry state ($u = 0$) where one layer is directly on top of the other, called AA stacking (Figure 1a center, 1b edges). Away from $u = 0$ are six energy minima, each one a different broken-symmetry ground state with an order parameter of magnitude $|u| = a$, where $a$ is graphene's bond length. These minima correspond to states that put one of the first layer's sublattice atoms (A or B) directly on top of its opposite sublattice atom (B or A) in the second layer, called AB or BA stacking, respectively, or collectively, called Bernal-stacking. Adjacent minima can be most easily traversed by the local change $\Delta u$ in the order parameter across an AB-to-BA stacking boundary. As shown in Figure 1a, these translations come in three types depending on the direction of $\Delta u$, which we label with colors red, green, or blue.

These AB and BA phases can be directly imaged using dark field transmission electron microscopy (DF-TEM).[7-8] An aperture in the diffraction plane of the electron microscope selects electrons scattered through a narrow range of diffraction angles, distinguishing between regions of different crystallographic symmetry.[9] Imaging through the [-1010] diffraction angles reveals the AB and BA stacking domains, whereas imaging using the [-2110] spots visualizes boundaries between stacking domains. Figures 1c-d show a graphene bilayer grown by chemical vapor deposition (CVD)[10] and supported by ~2 additional graphene sheets at 16° and 31° relative to the bilayer, imaged using these techniques. In Figure 1c, a striking hexagonal array of AB and BA domains is observed. The direction of the order parameter change, $\Delta u$, across each domain boundary is shown in the color-composite image in Fig 1d. Here, images from three of the [-2110] diffraction spots have been colored red, blue, and green, respectively, to match the translations shown in Fig 1a, and summed (see supplementary materials for more details).

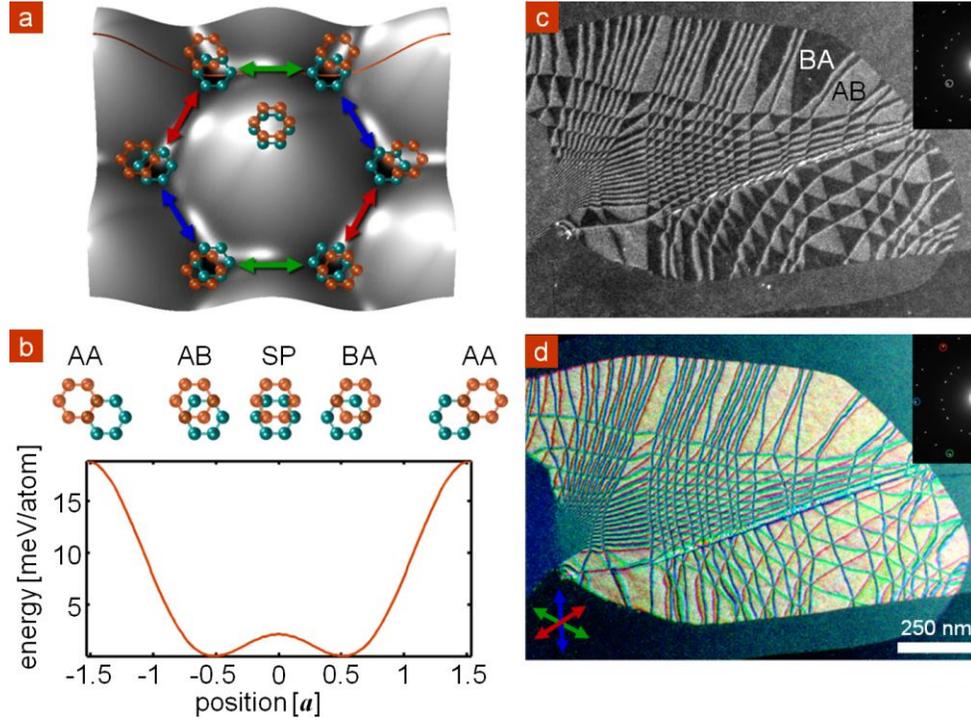

**Figure 1. a** van der Waals energy landscape for translating one graphene layer across another (adapted from ref. 6) with the corresponding orientations of the two layers shown schematically in orange and teal. The central location corresponds to AA stacking, having an order parameter vector, **u** = 0. Around this are six energy minima where |**u**| = $a$, corresponding to Bernal-stacked graphene. The two mirror-symmetric phases of Bernal-stacked graphene, AB and BA, are related to each other by three distinct low-energy translation directions |Δ**u**| = $a$ indicated by red, green, and blue arrows. **b** A horizontal line cut through the energy landscape in b, along an armchair direction, reveals that AB is connected to BA through a saddle-point (SP) having an energy of 2.1 meV/atom, a factor of ten lower than the energy of AA-stacked graphene. Across this cut, from left to right, the upper graphene sheet, shown in orange, translates to the right with respect to the lower sheet, shown in teal. **c,d** Dark-field TEM images of bilayer graphene, imaged through an aperture in the diffraction plane, as indicated by circles in the inset. The bilayer graphene is supported by ~2 additional graphene sheets at 16° and 31° relative to the bilayer, which are invisible when imaging though the selected diffraction angles. **c** At non-zero sample tilt, selecting electrons from the [-1010] family of diffraction angles enables us to distinguish AB (gray) from BA (black) domains. **d** Three DF TEM images taken from the [-2110] diffraction angles indicated in the inset, are overlaid in red, blue, and green. Imaged this way, each line is a AB-BA domain boundary, with its color indicating the armchair direction along which the relative translation between graphene layers occurs.

Using these images, we can immediately determine whether a boundary is a tensile strain boundary (Δ**u** perpendicular to the boundary), a shear strain boundary (Δ**u** parallel to the boundary) or somewhere between. These strains can be summed to obtain the global interlayer biaxial ($\bar{\nabla} \cdot \mathbf{u}$) and rotational ($\bar{\nabla} \times \mathbf{u}$) strain in the sample. In Figure 1d, most of the translation vectors are parallel to their boundaries, indicating shear; the observed pattern results from a global relative interlayer rotation between the two graphene sheets. Thus the observed triangular pattern (which has been previously observed in trilayer graphene[11] and graphite[12]) is similar to a Moiré pattern with the notable difference that, locally, the lattice has relaxed into commensurate Bernal-stacked phases of constant **u** separated by incommensurate domain walls, each associated with one of the three interlayer translation vectors Δ**u**.

Topological point defects are also evident in Figure 1. Figure 2a shows an enlarged region of the color-composite image from Figure 1d where three domain walls intersect. Superimposed in black are the inferred directions of the order parameter **u** for each domain, based on the transitions. The order parameter **u** rotates by 2π on a path that encloses the intersection point. This is thus a topological defect analogous to, e.g., a vortex in a superconductor. As with a superconducting vortex, the order parameter must vanish at the center, corresponding here to AA-stacked graphene (**u** = 0). A topologically equivalent Moiré structure is shown in Figure 2b; at the center of such structures, AA-stacked graphene (**u** = 0) is seen.

We use fifth-order aberration-corrected annular dark field (ADF) STEM to directly image the stacking with atomic resolution. An electron beam with a ~1.3 Å full-width at half maximum (FWHM) is scanned over the sample and the scattered electron intensity is recorded as a function of the beam position. Figure 2c shows the core of a topological defect where 6 domains meet, showing (bright) atoms in a hexagonal lattice, as is characteristic of AA-stacked bilayer. In AA-stacked graphene, all atoms in one layer are directly above those in the other, so each atom is visible, and all have similar brightness. The surrounding AB and BA domains appear considerably different. Figure 2d shows an example of Bernal-stacked bilayer domain, where we observe bright spots with hexagonal symmetry and a spacing of 0.25 nm, close to $\sqrt{3}a$. These spots correspond to the sites in Bernal-stacked bilayer where two atoms are stacked on top of one another; coherent scattering makes the intensity 3-4 times brighter than for individual atoms.[13] Thus, with AA at its center and AB and BA domains surrounding, the sixfold pattern in Figure 2a is a direct manifestation of the sixfold-degenerate energy-level landscape shown in Figure 1a. The AA cores are very high-energy, so they effectively pin the intersections of the three lines together. We rarely observe crossings of domain boundaries that do not respect this 3-fold rule.

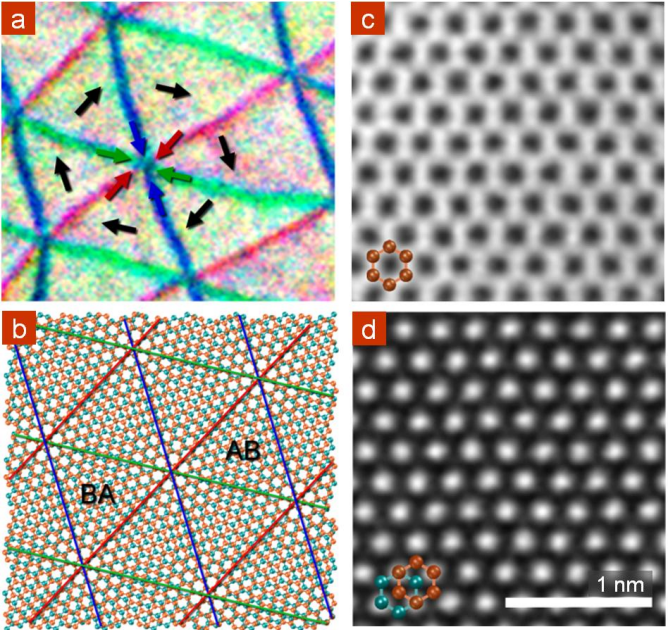

**Figure 2. a** An enlarged region of Figure 1d, showing a topological defect where six domains meet. Each domain (white) is associated with a different order parameter vector, **u** (black), and each boundary corresponds to an interlayer translation, **Δu**, as one traverses the boundary in the clockwise direction, shown in red, green and blue. **b** Schematic of two graphene sheets rotated relative to each other, showing a Moiré pattern that is topologically equivalent to the structure in Figure 2a. Alternating AB- and BA-like regions surround an AA-like core. **c** An atomic-resolution STEM image of the center of a region like that in Figure 2a where six Bernal-stacked domains meet, showing that such regions exhibit the energetically-costly AA-stacking. In AA-stacked graphene all atomic sites are visible in a hexagonal array, as indicated by the schematic. **d** A nearby Bernal-stacked region, for reference. In Bernal-stacked graphene, only half of the lattice sites are visible—those corresponding to atoms stacked directly on top of one another, as indicated by the schematic. In c and d, respectively, 3 and 7 frames were cross-correlated and averaged, after applying a 0.2 Å low-pass filter (far below our ~1.3 Å probe size), see supplemental materials for raw images and details.

We now examine the soliton boundaries between two stacking-phases with atomic resolution. STEM images of boundaries between AB and BA domains exhibiting concentrated shear and tensile strain are shown in Figure 3a and 3d, respectively. Towards the right and left sides of each image, we observe bright spots corresponding to Bernal-stacked bilayer. Towards the center of the boundary the brightness decreases, and this hexagonal pattern evolves into linear features that are horizontal for the shear boundary and vertical for the tensile boundary. This pattern results from the near-overlap of lines of zig-zag atoms that occurs as the two layers translate across each other vertically and horizontally, respectively, as indicated schematically in Figures 3c and 3f. Figures 3b and 3e display corresponding simulations of STEM images using Multislice quantum mechanical scattering calculations,[14,15] showing excellent agreement with the data (see supplementary materials for details).

The widths of the transition regions are different in the two cases. Figure 3g shows vertical line averages of images 3a and 3d, indicating that the shear boundary is significantly narrower than the tensile boundary. For the shear boundary, the average full-width at half-maximum (FWHM) is 6.2 ± 0.6 nm, while for the tensioned boundary the average FWHM is 10.1 ± 1.4 nm (see supplementary materials for details). These widths correspond to maximum strains in each layer of 0.8% and 0.5% for the shear and tensile boundaries, respectively, which occur at the center of each soliton. Figure 3h displays the soliton width (FWHM) versus the absolute value of the soliton angle $\phi$ (the angle between **Δu** and the boundary normal) obtained via STEM, as described in the supplementary materials. The soliton width varies with angle, having a maximum FWHM of ~11 nm at 0°, corresponding to purely tensile solitons, and decreasing to a minimum of ~6 nm at 90°, corresponding to purely shear solitons.

The observed widths can be understood as competition between strain energy in the transition region and the misalignment energy cost per unit length of the soliton: $E \cong \frac{1}{4}ka^2/w + V_{sp}w$, yielding an equilibrium width: $w_{eq} = \frac{a}{2}\sqrt{k/V_{sp}}$. Here, $k$ is the stiffness, and $V_{sp}$ is saddle-point energy per unit area in Figure 1b, and $a = 0.141$ nm is the bond-length in graphene. The Young's modulus, $Et = 340$ N/m,[16] is larger than the shear modulus $Gt \sim Et/(2(1+\upsilon)) = 142$ N/m, where $\upsilon$ is the Poisson ratio, predicting that the ratio of the widths of the tensile and shear boundaries is 1.5. This is in reasonable agreement with the experimental results in Figure 2. A more complete description is given by the two-chain Frenkel-Kontorova model[6], which predicts soliton boundaries between the domains to have width:

$$w_{eq} = \frac{a}{2}\sqrt{\frac{k}{V_{sp}}} = \frac{a}{2}\sqrt{\frac{1}{V_{sp}}\left(\frac{Et}{(1-\upsilon^2)}\cos^2\phi + Gt\sin^2\phi\right)}$$

where the boundary has an interlayer translation at an angle $\phi$ relative to the boundary normal. By relating $w_{eq}$ empirically to the FWHM through STEM image simulations (see supplementary materials), we fit the width-vs-angle model to the data, choosing to treat $Et$ as fixed at its measured value of 340 N/m,[16] and taking the Poisson ratio as that measured for graphite,[17] $\upsilon = 0.16$; we then use $V_{sp}$ and $Gt$ as fitting parameters, and overlay the result in Figure 3h (solid line). We obtain $V_{sp} = 1.2$ meV/atom, $Gt = 130$ N/m. These values are in excellent agreement with those predicted by theory—$V_{sp} = 1$-$2$ meV/atom,[6,18,19] $Gt = 142$ N/m, and imply a line tension for the domain walls in the ~100 pN range. We also performed width measurements on a greater number of solitons via DF TEM, observing similar qualitative behavior in the angular-dependence of width, but greater variability due to the influence of out-of-plane corrugations in the graphene on the DF TEM contrast across solitons, as shown and discussed in the supplementary materials.

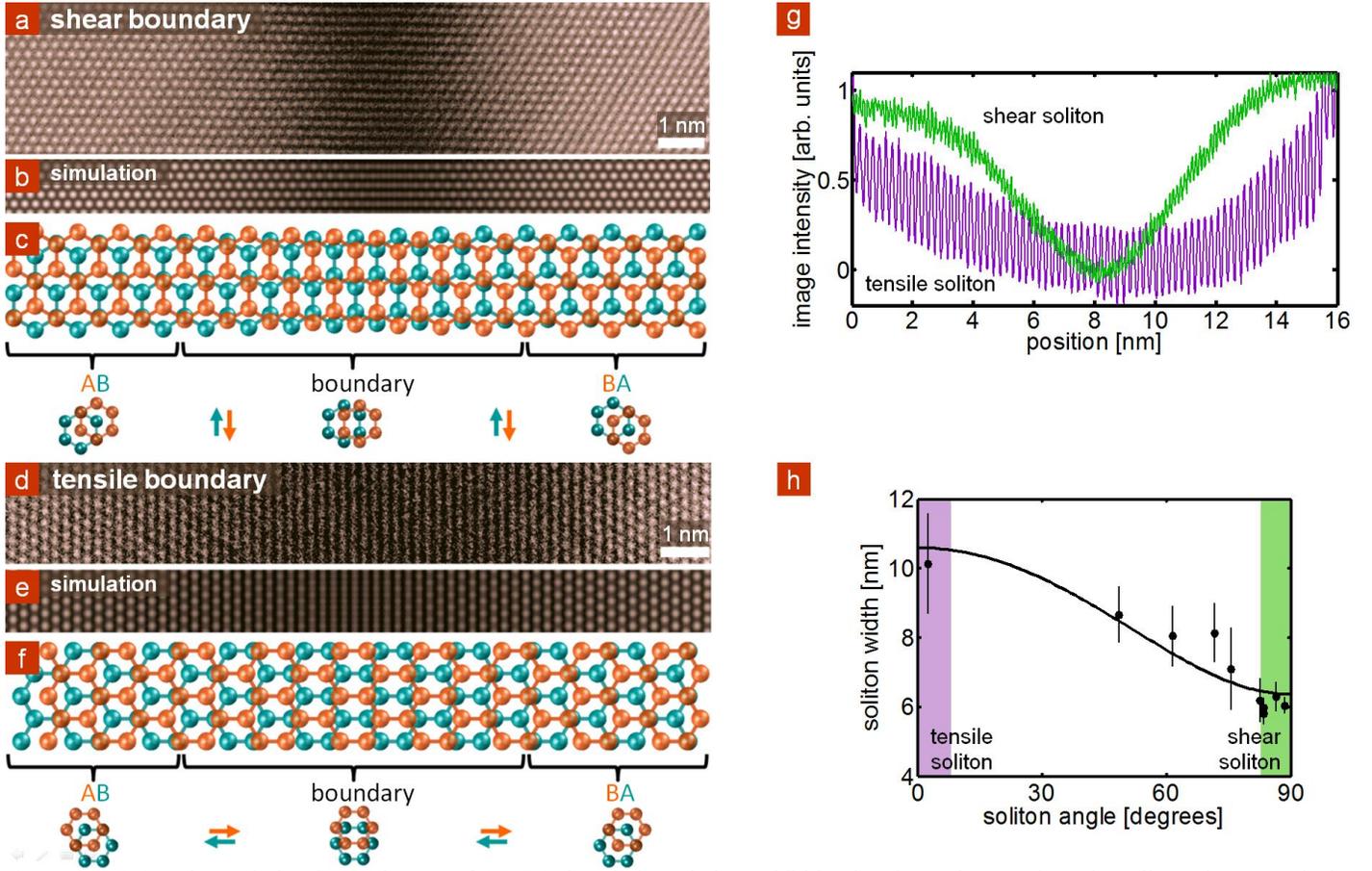

**Figure 3. a,d** Atomic-resolution STEM images of AB-BA domain boundaries, exhibiting interlayer shear strain and tensile strain, respectively. As one moves across the boundary from left to right, the two sheets translate relative to each other in opposite directions, as indicated by the schematics in c and f. Each image is an average of 4 adjacent regions along a boundary (details in supplementary materials). **b,e** Simulated STEM images of shear and tensile boundaries, respectively, show good agreement with the experimental images in a and d. The atomic coordinates have been specified by the soliton solution to the sine-Gordon equation,[6] discussed in supplementary. **c,f** Schematics showing shear and tensile boundaries, respectively (not to scale). In c, from left to right, the orange lattice translates downward, while the teal lattice translates upwards, completing a one-bond-length armchair-direction interlayer translation from AB to BA. Similarly, in f, the orange lattice translates to the right, while the teal lattice translates to the left. **g** Vertical line averages of the images in a (green) and d (purple) reveal that the STEM contrast profile across the boundary is approximately Gaussian, and that the shear boundary is significantly thinner than the tensile boundary. **h** Full-width at half-maximum (FWHM) for the STEM intensity profile for a few different solitons as a function of the absolute value of the angle between the interlayer-translation direction, and the soliton boundary-normal. The fit is given by the equation in the text, and indicates that the angular dependence of soliton width is explained by the decrease in stiffness associated with a change from tensile strain to shear strain.

Finally, we examine the mobility of the soliton boundaries. Figure 4a-c shows a series of DF TEM images taken with a large beam current (3.6x10$^4$ e$^-$/nm$^2$/s), at 80 keV. The boundaries fluctuate as shown in supplementary video, V1, shifting by tens of nm on the scale of minutes, as indicated by the arrows in Figure 4a-c. We also see instances where soliton boundaries undergo topological rearrangements. Figure 4d-e shows one such example, in which two solitons having opposite translation directions, $\Delta u$, appear to have contacted each other and annihilated. Since each soliton has an energy cost associated with it, which can be eliminated if this pair combines, these solitons are attracted to each other[20] and can annihilate via an interlayer translation of the intervening domain, here, labeled $u_0 + \Delta u$.

Although motion occurs at high beam currents, in general, at low beam currents (80keV, ~3x10$^3$ e$^-$/nm$^2$/s) and temperatures below 800 °C, motion is rare. Above 1000 °C, motion becomes more prevalent, as shown in supplementary videos, V2-5. The first and last frames in a temperature series from 1000 to 1200 °C are displayed in Figures 4f and g, showing that AB and BA domains anneal to form more regular structures, with shorter, straighter boundaries. In the videos, the boundaries typically move in discrete steps, which we attribute to pinning of the solitons by disorder and out-of-plane wrinkles. All of these images and videos show that these solitons are flexible and mobile, and that they can form ordered and complex patterns.

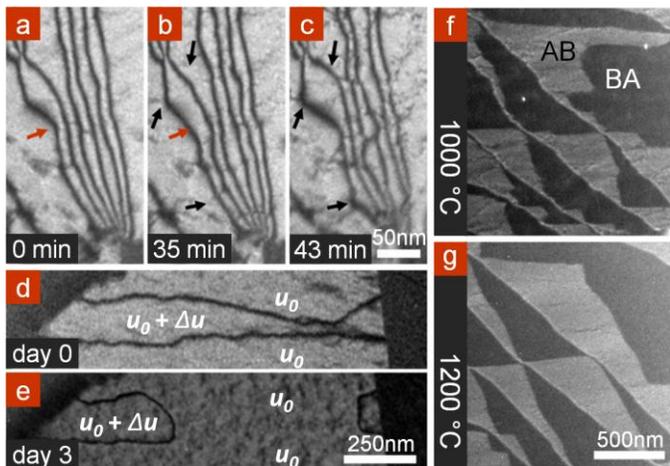

**Figure 4. a-c** [-2110] DF TEM images of interlayer solitons taken over the course 43 minutes while imaging at high beam intensity (see video V1 in supplementary materials). Under the influence of the beam, the soliton positions fluctuate by as much as 20 nm, as indicated by the arrows. **d-e** A pair of solitons having opposite translation directions (i.e. boundaries on either side of a region that is one-bond-length-shifted from the surrounding bilayer) may annihilate each other, as appears to have happened in the 3 days between which these images were taken. In the center, where, in d, the solitons are in close proximity, they have joined and disappeared in e. **f-g** First and last frames of DF TEM taken through a [-1010] diffraction angle at 1000 °C and later at 1200 °C, respectively. The boundaries between Bernal-stacking domains get shorter and straighter with time and temperature. For videos, see supplemental materials, V2-V5.

The complex and intriguing patterns of soliton boundaries observed here suggest that they will be an ideal laboratory for studying the physics of topologically-protected edge states. For example, the arrays of domain walls seen in Figure 1 are a new kind of superlattice structure that has only just begun to be studied theoretically[21] and may be relevant for recent measurements of the electronic structure of twisted bilayer graphene.[22-23] Furthermore, it may be possible to create devices where the motion of a single domain wall completely changes the conductance of a device in a manner analogous to magnetic domain wall magnetoresistive devices.[24-27] These solitons may also provide an explanation for the thermoelectric response at domain walls in multilayer graphene[28], as well as for the mystery of excess subgap transport typically seen in bilayer graphene transport experiments, where a perpendicular electric field is used to open a bandgap in AB or BA stacked graphene.[1-3,29] Recent theory predicts that a topologically protected 1D electronic state will form at the soliton boundary[4-5], and these 1D conducting pathways may be the major source of conduction in these samples. The properties of these 1D states depend on the width and orientation of the domain walls[4], which the measurements above provide explicitly.

## Methods

**Graphene Growth and Transfer** Large-grain (30-100μm) graphene was grown on copper foil (Alfa Aesar Cat#13382), by chemical-vapor deposition (CVD), using the enclosure method of Li *et al.*[10] using methane and hydrogen flow rates of 1-3 sccm, and 60-120 sccm, respectively, at 980 °C for 2 hours, then cooled. The resulting graphene is predominantly monolayer, with ~10 μm six-fold symmetric star-shaped bilayer and multilayer patches at many of its nucleation sites. We then use the methods of Huang *et al.*[9] to transfer the graphene to 200 nm nitride TEM grids (Ted Pella #21535-10), carbon grids (Quantifoil Q250-AR2), or heatable ceramic grids (Protochips E-AHF21).

**ADF-STEM** For STEM imaging, we used a NION ultra-STEM100, operated at 60 kV. Imaging conditions were similar to those used in References 9 and 30. Using a 25-mrad convergence angle, our probe size was close to 1.3 Å. The images presented in Figures 2 and 3 were acquired with a low-angle annular dark-field detector with acquisition times between 16 and 40 μs per pixel. Samples were baked for >10 hours at 130°C in ultra-high vacuum before loading into the microscope.

**DF-TEM** TEM imaging and diffraction were conducted using a FEI Technai T12 operated at 80 kV. Acquisition times for dark-field TEM images were 20 s per frame. We used displaced-aperture DF-TEM for the images in the main text. For in-situ heating, we used electrically-contacted pre-calibrated Protochips Thermal E-chips with Aduro sample holder, which allow heating up to 1200°C.

**Multislice image simulations** We simulated ADF-STEM images using numerical scattering calculations in E.J. Kirkland's multislice code. In this code, a full quantum mechanical multiple scattering simulation of electrons is propagated through multi-layered atomic membranes, producing quantitative simulations of dark field detector signals.[14] Atomic scattering factors are characterized by a 12-parameter fit of Gaussians and Lorentzians to relativistic Hartree-Fock calculations.[15]

**Acknowledgments** We thank Eun-Ah Kim for useful discussions. This work was supported by the Air Force Office of Scientific Research (AFOSR) through the Graphene MURI and individual grants (FA9550-09-1-0691 and FA9550-10-1-0410), and the National Science Foundation (NSF) through the Cornell Center for Materials Research (NSF DMR-1120296), which also provided electron microscopy facilities. This work was also partially funded by the SAIT GRO program and the National Research Foundation of Korea. Sample preparation was performed in part at the Cornell Nanoscale Science and Technology Facility, the Cornell node of the National Nanofabrication Infrastructure Network, funded by the NSF. P.Y.H. was supported under the National Science Foundation Graduate Research Fellowship Grant No. DGE-0707428. L.B. was partially supported by a Fullbright scholarship.

# Supplementary Materials

**DF TEM and composite "order parameter vector" images from [-2110] diffraction spots**

As stated in the text, if an aperture is placed in the diffraction plane at one of the angles corresponding to planes of atoms along the zigzag direction (the [-1010] family of diffraction angles) then at non-zero sample tilt, AB and BA are no longer symmetric with respect to the beam axis, and one phase appears bright while the other is dark.[7]

To image the AB/BA soliton *boundaries* on a few-micron scale, we apply the same technique, only instead of using the "inner" [-1010] diffraction spots, we use the aperture to select the "outer" diffraction spots—the [-2110] family of diffraction angles. For this family, in which electrons scatter from planes parallel to a given armchair direction, if a boundary translation, **Δ𝑢**, has a component perpendicular to that armchair direction, its contrast will change relative to the adjacent Bernal-stacked regions. Figure S1a-c displays a series of DF TEM images of the sample in Figure 1c-d. Figure S1f is taken directly from 1c, while in a-c, we have used an aperture to select three different [-2110] angles, indicated by the circles in Figure S1e. Comparing each of the boundary images, a-c, to the domain image, f, we notice that one third—and a different third—of the boundaries in each image, a-c, is invisible. From this we infer that the interlayer translation occurring across a given invisible boundary is precisely along the armchair direction that corresponds to the diffraction angle through which the image was taken. We indicate these boundaries schematically by dashed lines in Figure S1a-c,f. Thus each boundary represents a single-bond-length interlayer translation, in agreement with what we might expect by examining Figure 1a, where the minimal-energy path connecting AB to BA through a saddle-point corresponds to translation along one of three armchair directions.

Figure S1d shows a composite image in which we have colored the images from each of the three [-2110] diffraction spots, S1a-c, red, blue, and green, respectively, and merged them to create the image shown (in a manner identical to that used for Figures 1d, and 2a). In this image, each boundary has a color corresponding to the order parameter vector, **Δ𝑢**, indicated by arrows in Figures S1d, and 1d. Moving across the sample in Figure 1d from left to right, the density of boundaries decreases, from 1 per 6 nm, to 1 per 90 nm, corresponding to a decrease in relative global twist from 1.4° to 0.1°, which we posit occurred as a gradual interlayer rotation-relaxation process during the CVD growth from left to right.

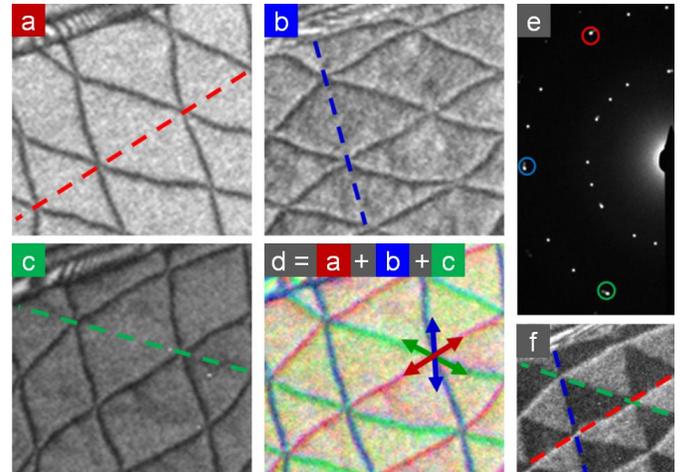

**Figure S1.** Composite-image construction. **a-c** DF TEM images taken through apertures in the diffraction plane, as indicated by the similarly-colored circles in Figure S1e. In each image two of the three domain boundaries are visible. The "missing" boundary in each (dashed lines) corresponds to a boundary with interlayer translation parallel to the diffraction planes being imaged. **d** Composite image constructed by coloring a-c red, blue, and green, respectively, and summing. **e** Diffraction image for this sample, showing the locations of the apertures used for imaging a-c. **f** DF TEM image of the sample in a-f taken through one of the "inner" [-1010] diffraction spots, indicating the locations of AB and BA domains.

**Linear global interlayer strain example**

The "rotational interlayer strain" sample in Figures 1 and 2, has a striking and easily interpreted structure, but is not the most common type of sample. Among tens-to-hundreds of samples imaged, we saw this sort of hexagonal/triangular pattern only 4 times. More typically, interlayer strain is less regular, often exhibiting some global interlayer uniaxial strain—likely related to the terraced structure in the copper growth substrate[7] (as we have also seen in unpublished work). An example of such a uniaxially-strained case is shown in Figure S2. Figure S2a is an "AB/BA domain" image, taken from a [-1010] diffraction angle, while S2b is a composite "boundary" image generated from the [-2110] family of diffraction angles as described for Figures S1d, 1d and 2a above, and in the main text.

In the case of this sample, the bilayer accumulates interlayer strain across the sample vertically. The translations having a significant shear component (red and green) largely cancel out, leaving an accumulation of strain, primarily due to the nearly pure-strain boundaries (blue). Two subtleties in this latter case are worth noting. Firstly, due to the energy landscape, a sample with large relative interlayer strain globally will always have some locally-sheared boundaries, since this is the only way to accumulate strain while avoiding a translation through an energetically-unfavorable AA stacking, see Figure 1a. Secondly, there are some interesting topological features in this sample which cannot be explained by the presence of interlayer strain and shear between two stacked sheets of *pristine* graphene, but instead arise from a topological point defects having non-zero in-plane Burgers vector.

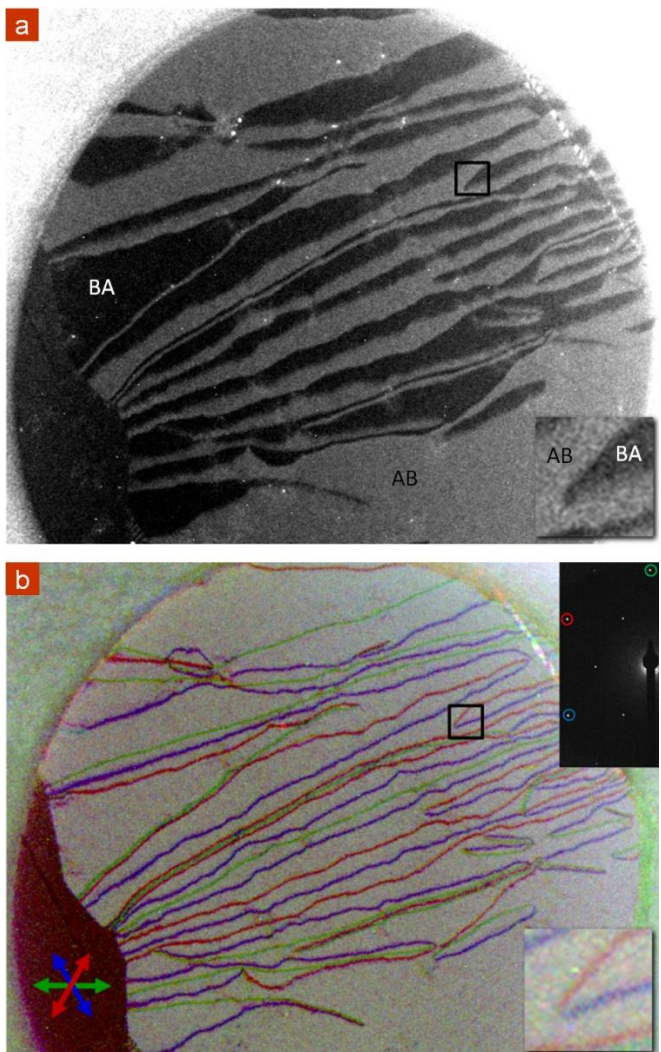

**Figure S2.** Example of a bilayer sample exhibiting predominantly linear global strain. **a** DF TEM image taken from one of the [-1010] diffraction angles showing lines of alternating AB- and BA-stacked graphene. **b** Composite DF TEM image taken from [-2110] angles, using the same methods as those used for Figure 1c, coloring the soliton boundaries according to their interlayer translation vectors, **Δu**, as indicated by the arrows. Insets highlight an interesting defect, as discussed in the supplemental text.

For the interested reader, one of these features is highlighted in the insets to Figure S2. Notice that two different translation vectors, **Δu** (red and blue lines), are associated with the boundary between a single AB- and single BA-stacked region. If the two layers in the bilayer were pristine graphene, the order parameter vector, **u**, we would assign to the BA region based on the known shift at the red boundary would be inconsistent with that assigned based on the blue boundary. One explanation for this apparent inconsistency, is that one of the two layers is missing a (zigzag) line of atoms, and has been stitched together with an offset that directly corresponds with the difference between the vectors associated with the red and blue domain boundaries. Or stated another way, there is a point defect at the intersection of the red and blue lines having a non-zero Burger's vector. To our knowledge this is the first time such a defect has been identified in graphene, and appears in many of our CVD-grown samples. In this image alone, there are more than 10 such defects. (For those interested in imaging such defects with atomic resolution, this DF TEM technique is useful for quickly identifying, to within ~10 nm, where to look. In attempting to image some of these defects, ourselves, we found that of ~10 that we tried to image with atomic resolution, all were covered with PMMA/etchant residue, perhaps due to increased reactivity.)

**Averaging and cross-correlating images for Figure 2**

In Figure 2c and 2d, respectively, 3 and 7 frames were cross-correlated and averaged, after applying a 0.2 Å low-pass filter. Figure S3a and S3b show examples of the raw images from which, respectively, 2c and 2d were taken. The cross-correlation was done using Matlab's image processing toolbox, in a two-pass registry. The first pass registered all images to the first frame in the stack, the second registered all images, including the first, to the average registered image from the first pass. After the second pass, the registered images were averaged, and the grayscale was adjusted to increase the contrast. The low-pass filter applied to the images was a standard Gaussian filter with $\sigma = 0.2$ Å.

**Simulated STEM images and soliton model**

To simulate the STEM images presented in Figure 3, we used E.J. Kirkland's Multislice code, as described in Methods in the main text. The atomic coordinates in the simulated image (and also in the schematics) were specified by using the two-chain Frenkel-Kontorova model[6] to describe the interlayer translation, $\Delta u$, in the boundary region in terms of the sine-Gordon equation:

$$\frac{ka^2}{4}\frac{\partial^2 \Delta u}{\partial x^2} = \frac{\pi}{2}V_{sp}\sin(2\pi\Delta u)$$

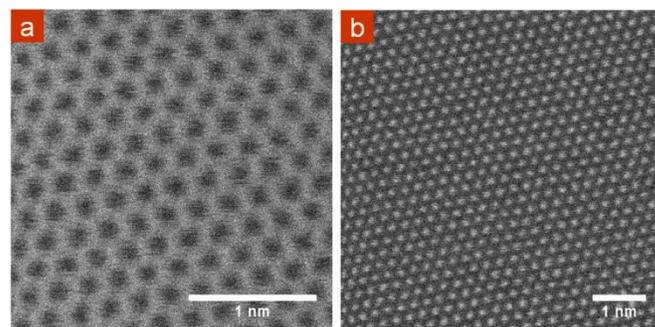

**Figure S3. a-b** Raw STEM images of AA- and AB-stacked graphene, respectively. Stacks of 3 and 7 similar images were cross-correlated and averaged and contrast-adjusted to generate Figures 2c-d.

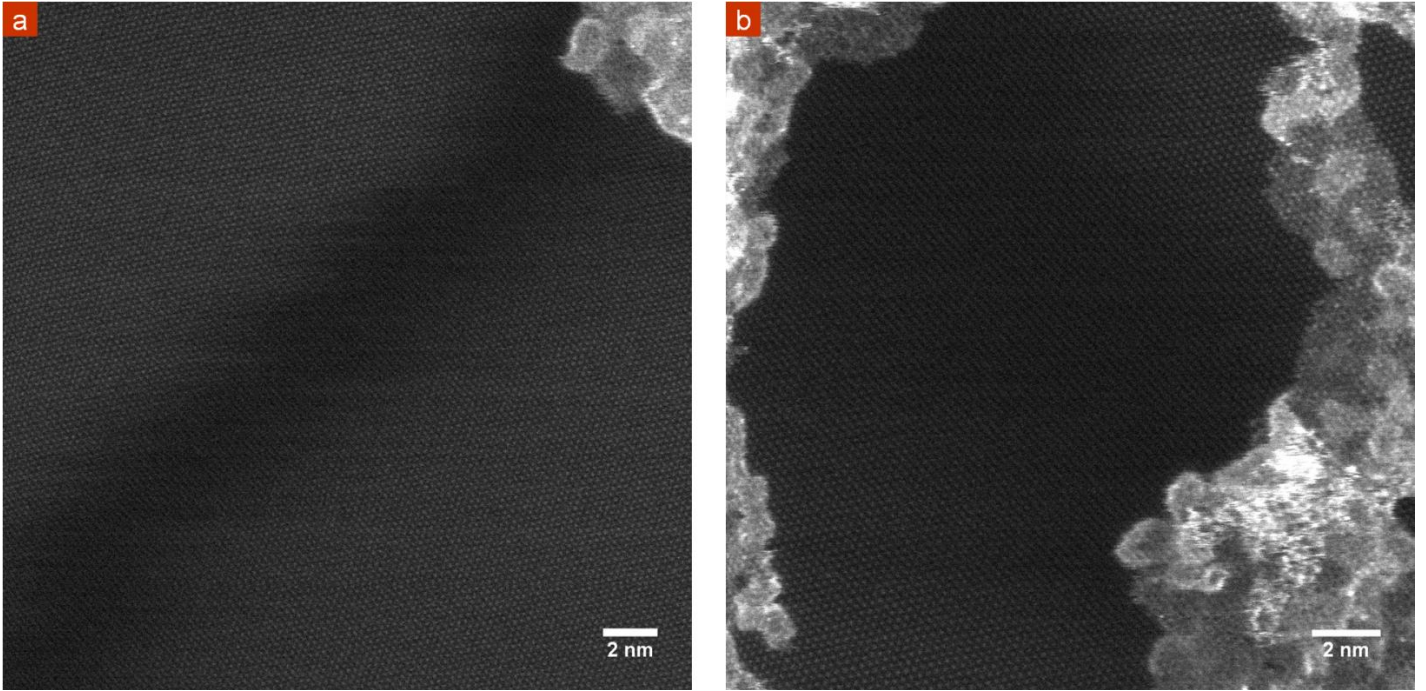

**Figure S4. a-b** Raw STEM images of shear- and tensile-strain soliton boundaries. 3-4 adjacent regions along these solitons were averaged and contrast-adjusted to generate Figures 3a and 3d.

Here, $k$ is the stiffness, and $V_{sp}$ is saddle-point energy in Figure 1a, $a = 0.141$ nm is the bond-length in graphene, and $\Delta u$ is the broken symmetry order parameter, which varies from 0 to 1 across the boundary region. The first term is elastic energy stored in the boundary region, and the final term is the misalignment cost associated with non-AB/BA stacking.

This equation has soliton "kink" and "anti-kink" solutions of the form:

$$\Delta u_{\pm} = \frac{1}{2} \mp \frac{1}{2} \pm \frac{2}{\pi}\arctan\left(\exp\left(\frac{2\pi}{a}\sqrt{\frac{V_{sp}}{k}}x\right)\right)$$

The equilibrium width, $w_{eq} = \frac{a}{2}\sqrt{k/V_{sp}}$ was used as a fitting parameter to match the FWHMs of the simulated images with those of the STEM images in Figure 3. Upon obtaining the Multislice output, a Gaussian low-pass filter ($\sigma = 0.04$ nm) was applied to the simulated image to represent the incoherent probe size, again choosing this value based on a match with the STEM images.

**STEM FWHM**

In order to improve the signal-to-noise for our atomic resolution images, the composite images in Figure 3 were generated by averaging 3-4 regions in a single image that were adjacent to each other along the soliton. The raw images are shown in Figure S4a and S4b for Figure 3a and 3d respectively. The fits to the composite images in Figure 3 yielded FWHM of 13 and 5.9, for strain and shear, respectively. However, due to small motions of the soliton during imaging and slight in-plane curvature in the soliton, this averaging procedure leads to an apparent broadening of the soliton's width.

To avoid such broadening when determining the widths for Figure 3 and thus parameters of the soliton model (and also the cited "average FWHM" for shear and strain boundaries), we employ a second procedure for all STEM soliton width measurements. We fit Gaussians to line-cuts in the raw images parallel to the scan direction—averaging every 2-10 lines, depending on the size of the image—which eliminates the majority of the broadening due to fluctuations or curvature seen in the above averaging procedure. For S4, these cuts were in the horizontal direction. For the Gaussian fits, we used the center, $\mu$, height, $A$, and width, $\sigma$, as fitting parameters, and fixed the base of the Gaussian to be the average intensity of a region as far from the boundary as possible within the same image. Fits either having $\mu$ within 1 $\sigma$ of the edge of the image, or having a larger-than-median root-mean-squared-error (RMSE) were discarded. The resulting FWHMs from these fits were multiplied by the cosine of the angle between the scan direction and the boundary to obtain the boundary FWHMs. The error bars in Figure 3h are ± 1 standard deviation in the fitted width of a given soliton across all scan lines.

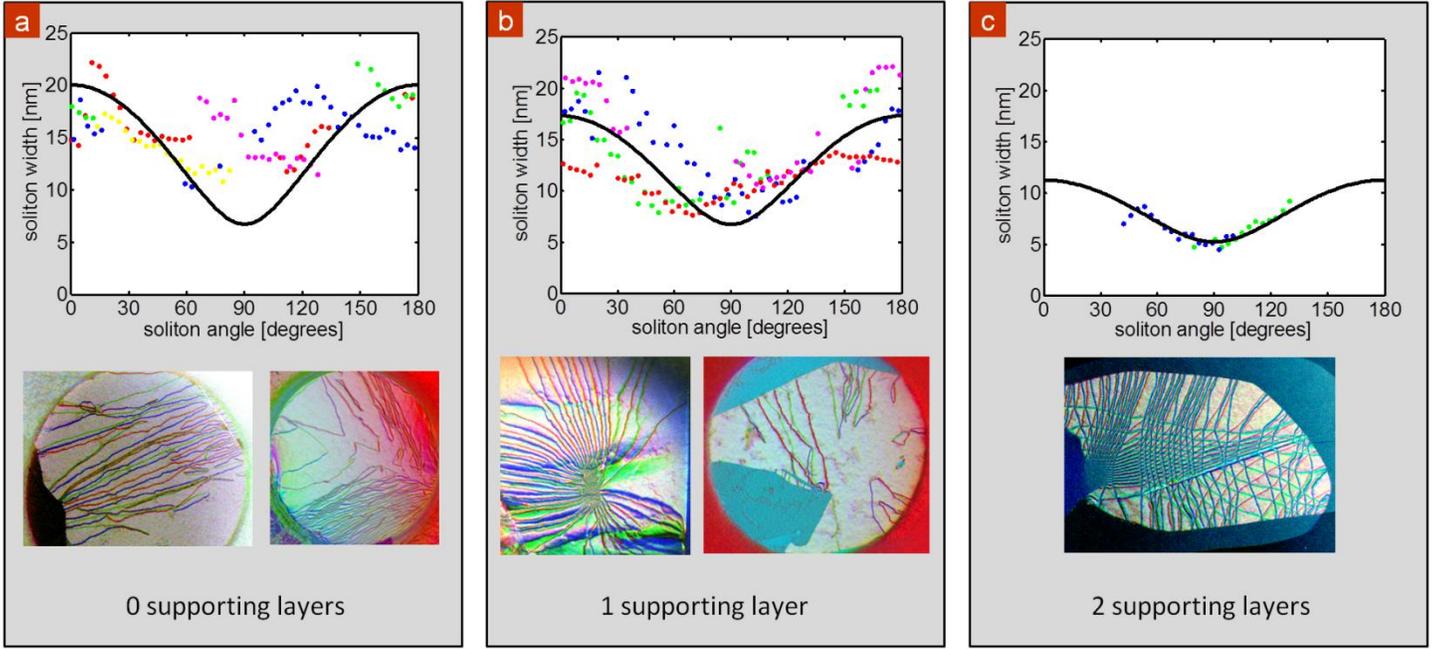

**Figure S5.** DF TEM images of soliton width as a function of angle, with corresponding composite images. For each sample, width measurements from two of the three soliton translation directions, $\Delta u$, are shown. **a** Samples freely-suspended bilayer graphene show considerable variability in the measured soliton width, presumably due to out-of plane corrugations in the graphene. **b** Samples with one additional graphene layer (at a non-Bernal-stacking angle) show qualitative agreement with our model, but considerable variability. **c** Samples with two additional supporting graphene layers show excellent agreement with our model and with the STEM width measurements.

**Relating STEM FWHM to Sine-Gordon**

To determine the soliton parameters based on our FWHM measurements we needed a way to relate the soliton width parameter to the FWHM. We accomplished this by fitting a polynomial function of $\Delta u$ (the change in the order parameter) to the average intensity across the multislice simulated STEM images in Figure 3. Since the coordinates of the atoms were generated using the solution to the soliton equation—see above—this enables us to fit a polynomial in $\Delta u$ which we can use to relate FWHM to soliton the width. For $\Delta u$ between 0 and 1, we find that the following polynomial fitted the multislice image well—and indeed, had lower RMSE than a Gaussian fit:

$$I(z) = Az^3 + Bz^2 + Cz + D$$

Here, $z = (\Delta u - \frac{1}{2})^2$, and $A$-$D$ are fitted parameters, having respective values: -25.8169, 4.7742, 1.7676, 0.6628. Fitting a Gaussian to this function for a few values of the soliton width (recall that $\Delta u$ is a function of the soliton width) allows us to establish a linear relationship between FWHM and soliton width:

$$w_{soliton} \equiv \frac{a}{2}\sqrt{\frac{k}{V_{sp}}} = A_1 w_{FWHM} + A_0$$

The fitted values for $A_1$ and $A_0$ are respectively: 1.458, 0.099. Using this relationship enables us to extract physical constants from our STEM measurements of the soliton's FWHM.

**DF-TEM width vs angle**

To image soliton boundaries in a larger number of samples, and on a ~micron length-scale, we use DF-TEM. We find very little preference for any one angle over the others, with many samples exhibiting boundaries at all angles. Figure S5 displays the boundary width versus angle $\phi$ obtained via DF-TEM. As was seen for the STEM measurements in Figure 3h, the soliton width varies approximately sinusoidally with angle, having a maximum FWHM at 0° (and 180°), corresponding to purely tensile solitons, and decreasing to a minimum at 90°, corresponding to purely shear solitons. The solitons appear wider than those measured by STEM, and have greater variability. This is likely the result of variations in the corrugations and built-in strain in the samples—to which width measurements performed using DF-TEM are more susceptible than those using STEM, where corrugated samples can easily be identified and rejected. In particular, corrugations parallel to a tensile boundary are expected to decrease the equilibrium width of the boundary, while increasing its measured width, due to out-of-plane-tilted bilayer being difficult to distinguish from interlayer-translated bilayer for small angles/translations. We find that as the number of supporting graphene layers—i.e. graphene layers oriented at some angle (>2°) with respect to the bilayer—increases from 0 to 2, shown in Figure S5a-c, respectively, the measured FWHM and the variability in FWHM measurements is reduced. This supports the view that corrugations are responsible for the variability in and broadening of measured soliton width, since the increasing stiffness associated with an increasing number of supporting layers reduces the amplitude of corrugations. With two supporting layers, the model fits well, and the measured strain soliton-FWHM is ~11 nm and the measured tensile soliton-FWHM is ~6 nm, in excellent agreement with our STEM measurements.

**Relating DF-TEM FWHM to Sine-Gordon**

The DF-TEM fits to the soliton width-vs-angle model are treated similarly to those for STEM. In DF TEM, the intensity collected through a [-2110] diffraction spot, at normal incidence, relates to the interlayer translation as:
$$I(\Delta u) \propto \cos^2(\Delta u)$$
Since the resolution of this technique is significantly below that of STEM, we must take into account the broadening of a soliton by its convolution with the finite-sized electron beam. In the case of resolution-broadening, the soliton FWHM will be given by:
$$w_{FWHM} = 2\sqrt{2\log 2 (\sigma^2_{measured} - \sigma^2_{resolution})}$$
We determine the resolution, $\sigma_{resolution}$, by, for each image, measuring the resolution-broadening of a graphene edge (often a bilayer-monolayer step), which we assume to be atomically sharp. We treat the image of the edge as the convolution between a Heaviside function and a Gaussian probe, and extract the width parameter, $\sigma$, for such a probe.

We automate the finding and fitting of the solitons in our outer-diffraction spot DF TEM images. Our algorithm first finds the boundaries, primarily by applying a threshold to the image, and assuming all pixels darker than a given threshold are soliton pixels. We then determine the orientation of the boundary by finding the ~20x20 px mask that minimizes the sum of squares between the image and mask, where the masks consist of a dark line drawn at some angle on a light background. We throw out error-prone regions (such as regions where two solitons intersect). We then fit a Gaussian at each soliton pixel, in a direction perpendicular to the soliton, averaging over the adjacent 3 pixels on either side, parallel to the soliton. Since some of the found pixels are not in fact solitons, and result in Gaussian fits with extremely large sigma (i.e. a flat region), we use the median width at each angle (rather than the mean) so as not to be strongly affected by such outliers.

Finally, we use the linear empirical relationship between the FWHM of $I(\Delta u)$ and the Sine-Gordon width to relate the fitted Gaussians to the sine-Gordon width.

**DF TEM videos**

**V1**: DF TEM video taken from the [-2110] family of diffraction angles, showing interlayer solitons fluctuating over the course 43 minutes under the influence of a high-intensity electron beam (3.6x10$^4$ e$^-$/nm$^2$/s, 80 keV). Each frame in V1 is an average of three images, each taken with a 20s exposure.

**V2-4**: DF TEM videos taken at 1000, 1100, 1200 °C, under low beam intensity (80keV, ~3x10$^3$ e$^-$/nm$^2$/s), using an aperture to select electrons from the [-1010] family of diffraction spots, showing AB and BA domains growing and shrinking as the solitons move. At our temporal resolution, motion often appears to occur in discrete steps. Upon first heating the sample, motion was significant at 1000 °C, as in V2. After heating to 1200 °C, cooling, and reheating to 1000 °C motion was negligible (not shown), suggesting that the initial motion at 1000 °C is primarily due to stress-relaxation. Videos were cross-correlated to remove sample drift, and each video frame is an average of five 20s exposures. V2-4 were taken over the course of 35, 27, and 27 min, respectively. (The isolated white pixels are dead pixels in the CCD, that appear to move due to the cross-correlation-based sample-drift correction.)

**V5**: DF TEM video taken at 1200 °C over the course of 138 min using the same imaging conditions and averaging procedure as V4. The sample has been tilted, leading to contrast among the domains and boundaries that appears different than that of the previous videos, V2-4.